\documentstyle[11pt,newpasp,twoside,epsf]{article}

\def\solar{\ifmmode_{\mathord\odot}\else$_{\mathord\odot}$\fi} 
\def\kms{km\thinspace s$^{-1}$}     
\def\deg{\ifmmode^\circ\else$^\circ$\fi}  
\def\arcs{\ifmmode {'' }\else $'' $\fi}  
\def\arcm{\ifmmode {' }\else $' $\fi}    
\def\mstar{M$_{HI_\ast}$}
\def\lya{Ly-$\alpha$} 

\begin{document}

\title{The Contribution of HI-rich Galaxies to the Damped Absorber Population
at $z=0$}

\author{Jessica L. Rosenberg \& Stephen E. Schneider}
\affil{Center for Astrophysics \& Space Astronomy, Department of Astrophysical
and Planetary Sciences, University of Colorado, Boulder, CO 80309}
\affil{Department of Astronomy, University of Massachusetts, Amherst, MA 01003}

\begin{abstract}
We present a study of HI-rich galaxies in the local universe selected from blind
emission-line surveys. These galaxies represent the emission-line counterparts of 
local damped Lyman-$\alpha$ systems. We find that the HI cross-section of
galaxies is drawn from a large range of galaxy masses below \mstar,  
66\% of the area comes from galaxies in the range 8.5 $<$ Log \mstar\ $ < $ 9.7.
Both because of the 
low mass galaxy contribution, and because of the range of galaxy types and
luminosities at any given HI mass, the galaxies contributing to the HI
cross-section are not exclusively L$_\ast$ spirals, as is often expected. The
optical and near infrared counterparts of these galaxies cover a range of types 
(from spirals to irregulars), luminosities (from L$_\ast$ to $<$0.01 L$_{\ast}$), 
and surface brightnesses. The range of optical and near infrared properties as 
well as the kinematics for this population are consistent with the properties 
for the low-$z$ damped \lya\ absorbers. We also show that the number of HI-rich 
galaxies in the local universe does not preclude evolution of the low-$z$ damped 
absorber population, but it is consistent with no evolution.
\end{abstract}

\section{INTRODUCTION}

The damped Lyman-$\alpha$ absorption-line systems (DLAs), absorbers with
N$_{HI} \ge 2 \times 10^{20}$ cm$^{-2}$, are often assumed to be the disks of
large spiral galaxies or their progenitors. There are not many of these systems
known at low-$z$: only 9 DLAs are know with $z < 0.5$ 
(Rao \& Turnshek 2000, Bowen et al. 2001, Steidel et al. 1994, Lanzetta et al. 
1997, and Le Brun et al. 1997), and only 2 of those are at $z<0.1$ where 
detailed observations are possible. Absorption line studies do not provide 
adequate statistical samples because their pencil beams survey only small 
volumes of space.

Low-$z$ DLAs have provided a few surprises. In particular, many of the 
optical counterparts are dwarf or low surface brightness galaxies rather than
bright spirals. The optical characteristics, as well as the metallicities of 
DLAs have lead to speculation that the absorption line population is biased 
against detecting the expected population of bright spiral galaxies. However, an
unbiased low redshift sample has never been constructed for study, which may
account for some of the surprising characteristics. We make use of blind 21 cm
surveys to construct a low redshift sample unbiased by a galaxy's optical
properties and use it to determine what we expect to find in the absorber
population.

What constraints does the number density of HI-rich galaxies in the local universe 
place on the low redshift evolution of DLAs (\S 3.4)? Are the kinematic 
properties of these $z$=0 absorbers consistent with those of the high-$z$ 
 population (\S 3.5)?  What are the optical 
properties of these $z$=0 galaxies and are they consistent with the properties 
found for nearby absorbers (\S 3.6)? 

\section{THE DATA}

For this study, we use the Arecibo Dual-Beam Survey (ADBS, Rosenberg \& 
Schneider 2000), a 21 cm survey that covered $\sim$ 430 deg$^2$ in the Arecibo 
main beam. The velocity coverage of the survey was --654 to 7977 \kms\ with an 
average $rms$ sensitivity of 3.5 mJy. The ADBS identified 265 galaxies, 
only a third of which had previously identified optical counterparts.The problem 
with comparing emission and absorption line studies are the differences in spatial 
resolution (0.25\arcsec-1\arcsec\ for absorption and $\sim$ 40\arcsec\ for VLA 
D-array studies).

All of the ADBS sources were followed-up at Arecibo or the VLA. We use the 99 VLA 
D-array maps to investigate HI size of HI-rich galaxies. The covering area of is 
determined in an isophote at a column density of 2$\times$
10$^{20}$ cm$^{-2}$. Because of the low resolution of these data, we have used 
isophotal fits rather than pixel-by-pixel measurements. However, the galaxies that
were mapped were selected to be the lowest mass sources in order to constrain the 
low mass end of the HI mass function. To fill in the sizes of high HI 
mass sources, we use information available in the literature (Martin 1998). These
data include a compilation of galaxy sizes at given column densities. We restrict
the use of these galaxies to the source for which there is a VLA, Westerbork, or
ATCA measurement at 
Log $N_{HI} = 20.3$ or for which the ratio of the  standard deviation to the HI
size at Log $N_{HI} = 20.3$ is less than 5\%. 

In addition to comparing the gaseous properties of the samples, we compare the 
stellar properties of HI-rich galaxies to those of the low redshift DLAs that 
have been found. To examine the stellar properties of the galaxies, we use
the 2 Micron All-Sky Survey (2MASS, Jarrett et al. 2000). 2MASS used 2 identical
1.3m telescopes in Tucson and Chile which provide simultaneous J, H, and K$_s$ 
observations. The observing time per point is only 7.8 seconds, so this survey is 
not ideal for finding faint and low surface brightness galaxies, but it 
provides a near infrared measurement for the brighter galaxies in our sample. 

\section{RESULTS}

\subsection{The HI Mass Function}

To probe the distribution of properties of HI-rich galaxies in the local 
universe, we must determine the number density (the HI
mass function) of these galaxies. Knowing the number density of HI-rich galaxies
locally, we can compare their properties to those of DLAs.

\begin{figure}[ht]
\begin{center}
\epsfxsize=2.5in
\leavevmode
\epsffile{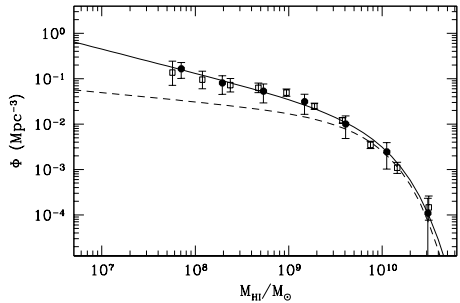}
\end{center}
\caption{The HI mass function as derived for sources in the ADBS (Rosenberg \&
Schneider 2001). The mass function was computed 2 ways: the open squares show
the results from the 1/{\it V$_{tot}$} method; the filled circles show the
result from the step-wise maximum likelihood method. The solid line is a
Schechter function fit to these data with a low-mass end slope of $\alpha$ =
-1.5. The dashed line is a Schechter function with $\alpha$ = -1.2 as found by
Zwaan et al. for their sample (1997) and as we found for the shape of the HI
mass function in the Virgo Cluster.}
\label{fig:mf}
\end{figure}

The derivation of the HI mass function depends on a careful assessment of the 
HI sensitivity function for the survey. A complete description of the HI mass 
function derivation for the ADBS can be found in Rosenberg \& Schneider (2001).
Figure \ref{fig:mf} shows the ADBS HI mass function. We have fit a Schechter
function to the results and find the faint-end slope of $\alpha$ = -1.5, the 
knee of the curve is at Log(M$_\ast$ /M$_\odot$) = 9.88, and the normalization is
$\Phi$ = 0.0058 Mpc$^{-3}$. Our result indicate that the population of small
HI-rich galaxies is larger than the number of faint galaxies found in optical
luminosity function studies. 

\subsection{Galaxy Sizes at 21 cm}

\begin{figure}[ht]
\begin{center}
\epsfxsize=2.5in
\leavevmode
\epsffile{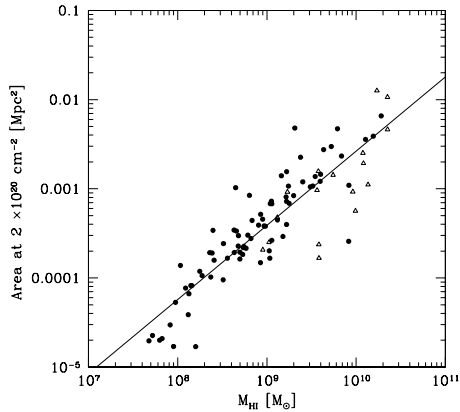}
\end{center}
\caption{The relationship between de-projected area calculated at 2 $\times 
10^{20}$ cm$^{-2}$ and HI mass for HI-rich galaxies. The filled circles are 
the points from the ADBS, the open triangles are from HI maps
published in the literature and compiled by Martin (1998).}
\label{fig:massarea}
\end{figure}

Figure \ref{fig:massarea} shows the relationship between de-projected HI area and
HI mass. Giovanelli \& Haynes (1983) found that the relationship between HI mass
and optical size could be used as a measure of HI deficiency in galaxies because
it was generally consistent among undisturbed galaxies. The 
correlation between HI mass and HI size, tighter than for optical size,
indicates a narrow range of average HI column densities. The outlying galaxies in 
Figure \ref{fig:massarea} show indications of being disturbed by interactions. For 
this study we scale the area by 0.64, the average projection factor for the 
sample.

The low resolution of the data and the use of global galaxy relations to derive
galaxy size relationships ignores the small scale properties of these galaxies. 
If the surface filling factor of  high column density HI (N$_{HI} > 2 \times 
10^{20}$ cm$^{-2}$) is small, we have overestimated the size of the damped region 
for each of these galaxies. It is useful to note, however, that there can also be 
regions of damped emission beyond our measured HI isophote which will contribute 
to a larger HI covering area. 

Detailed studies of the HI covering fraction have been done in very
few galaxies to date and have shown disparities in the covering fractions derived 
for different galaxies. Braun \& Walterbos (1992) find that the HI volume filling 
fraction in the Galaxy is $\sim$ 16\% while it is $\sim$ 38\% in M31. However, 
the surface filling fraction, the important value here, approaches 1 where the
emission brightness temperature exceeds $\sim$5 K (equivalent to a column
density of 2 $\times 10^{20}$ cm$^{-2}$ for an asymptotic temperature T$_\infty =
125$ K and a velocity width of 21.5 \kms). These numbers imply that the
correction for the covering fraction in these galaxies should be relatively
small.
	
\subsection{Galaxy Covering Fraction}

The relationship between HI mass and HI size, discussed in the previous
section, can be used the relationship to calculate the number of damped systems per
unit redshift as a function of HI mass. This function is the multiplication of the 
HI mass function and the HI mass versus HI size relationship and is given by:

\begin{equation}
dN(M_{HI})/dz = \Phi (M_{HI}) [Mpc^{-3}] \cdot A(M_{HI}) [Mpc^2] \cdot (c/H_0)
\end{equation}

where $\Phi (M_{HI})$ is the number density of galaxies in the mass bin as given
by the HI mass function and $A(M_{HI})$ is the average covering area in the
mass bin. We use H$_0$ = 75 \kms Mpc$^{-3}$ throughout, but note that dN/dz is 
independent of H$_0$.

\begin{figure}[ht]
\begin{center}
\epsfxsize=2.5in
\leavevmode
\epsffile{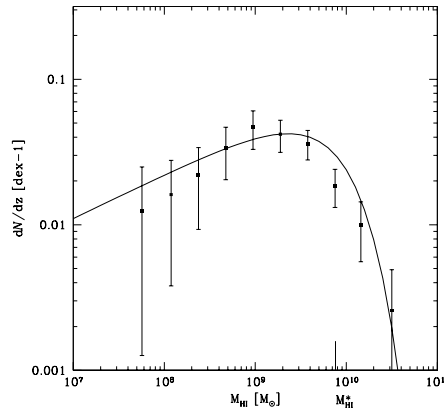}
\end{center}
\caption{The value of dN/dz as a function of mass for galaxies identified in 
the ADBS. 66\% of the area comes from galaxies in the range 8.5 $<$ Log
\mstar\ $ < $ 9.7.}
\label{fig:avgarea}
\end{figure}

Figure \ref{fig:avgarea} shows the resulting dN/dz versus HI mass function. This
figure shows that 66\% of the area comes from galaxies in the range 8.5 $<$ Log
\mstar\ $ < $ 9.7. The highest mass galaxies (Log \mstar\ $>$ 10.26) contribute
13\% of the area while the lowest mass galaxies (Log \mstar $<$ 8.5)
contribute 21\%. The cross-section of galaxies is not dominated by the highest
HI-mass galaxies, but is instead a contribution from a range of galaxy masses
below \mstar.

\subsection{Limits on the DLA Population in the Local Universe}

\begin{figure}[ht]
\begin{center}
\epsfxsize=2.5in
\leavevmode
\epsffile{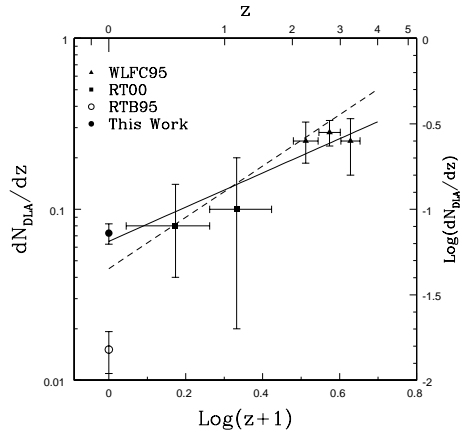}
\end{center}
\caption{dN/dz of the DLA population as a function of
redshift. The triangles are from Wolfe et al. (WLFC95; 1995), the squares are
from Rao \& Turnshek (RT95; 1995), the open circle is the $z = 0$ point for 
bright spiral galaxies (RTB95; 1995), and the
filled circle is the $z = 0$ point determined from the ADBS data in this work.}
\label{fig:dndz}
\end{figure}

Whether the DLA population evolves at low-$z$ remains difficult to answer
because of the poor statistics. The work of Rao \& Turnshek
(2000) used HI 21 cm Arecibo data from Rao et al. (1995) on 30 bright spiral 
galaxies 
to calculate dN/dz for damped systems in the local universe. Although the errors
are large, the data indicate that there has been strong evolution between $z$=2.0 
and $z$=0.5. However, the survey of large, bright 
spiral galaxies may not be representative of the HI-rich galaxy 
population. Figure \ref{fig:dndz} shows the same plot with the Rao et al. (1995) 
bright spirals (open circle) and the ADBS dN/dz value (filled
circle). Note that both $z = 0$ points suffer from measuring the gas in a large
beam, although the resolution is significantly better for the ADBS survey data.

Figure \ref{fig:dndz} shows dN/dz of the DLA population as a function of
redshift. The triangles are from Wolfe et al. (WLFC95; 1995), the squares are
from Rao \& Turnshek (RT00; 2000), the open circle is the $z = 0$ points
from the Rao et al. bright spiral galaxies (RTB95; 1995), and the
filled circle is the $z = 0$ point determined from the ADBS data in this work. 
The ADBS data is consistent with the q$_0 = 0$ no evolution model shown by 
the solid line in this figure. The
dashed line shows the $\gamma$ = 1.5 fit from Rao \& Turnshek (2000). The large 
error bars on these data and the uncertainty in the HI covering fraction means
that evolution of the DLA population at low-$z$ can not be ruled out. However, 
the known population of HI-rich galaxies is consistent with no evolution of the 
absorbers.

\subsection{Line Width Distribution}

Prochaska and Wolfe have analyzed the kinematics of 35 DLAs and used them to
constrain models of the origin of these sytems (Prochaska \& Wolfe 1997;
Prochaska \& Wolfe 1998; Wolfe \& Prochaska 2000). They assume that the galaxy
population is homogenous and therefore require massive spiral galaxies to
account for the largest observed linewidth sources in their sample. 
Additionally, they 
find that a population of spherical systems, kinematically similar to dwarf 
galaxies/protogalaxies, can be ruled out as the origin of the kinematics at
greater than 99.9\% confidence.

\begin{figure}[ht]
\begin{center}
\epsfxsize=2.5in
\leavevmode
\epsffile{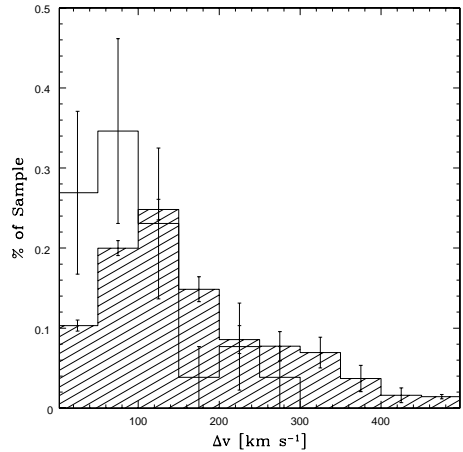}
\end{center}
\caption{A comparison of the kinematics of the DLAs, open histogram (Prochaska
\& Wolfe 1998, 2000) and of the ADBS galaxies (shaded histogram). The ADBS line
widths are consistent with the large linewdiths seen in the DLA sample.}
\label{fig:kinemat}
\end{figure}

Figure \ref{fig:kinemat} shows the percentage of the DLA sample (open
histogram) and of the ADBS sample (shaded histogram) with a given line
width. The error bars reflect the Poisson statistics in each bin. There are a 
total of 26 galaxies in the DLA sample and 265 in the ADBS. The ADBS data show 
that the population which dominates the HI cross-section is kinematically
consistent with the DLAs. There is a deficit of small line width sources in the
ADBS data relative to the DLA data. There are several possible explanations of 
this deficit: (1) the absorption line studies are biased towards lower line
widths because, with their higher spatial resolution, they tend to sample a 
smaller portion of the rotation curve; (2) the DLA data has not
been corrected for completeness so linewidth effects, such as the
lower contrast between continuum and absorption-line for higher velocity width
sources, may preferentially remove large line-width sources from the sample; (3) 
the most massive 
spiral systems may obscure the background AGN and therefore be missed in 
absorption, as has been discussed by other authors.

\subsection{Optical and Near Infrared Properties of Absorber Population}

Because L$_\ast$ galaxies dominate the light in the universe, and
M$_{HI_\ast}$ galaxies dominate the HI mass, it was generally
expected that the optical counterparts to the DLAs would be HI-rich L$_\ast$ 
galaxies - bright spirals. Instead, many of the low-$z$ DLAs are low 
luminosity and/or low surface brightness. Part of the explination for the
stellar properties of the DLAs is that there is a large dispersion in the
HI-mass optical luminosity correlation. Additionally, Figure \ref{fig:dndz}
shows that there is a large range of masses below M$_\ast$ that contribute to
the HI cross-section.

\begin{figure}[ht]
\begin{center}
\epsfxsize=2.5in
\leavevmode
\epsffile{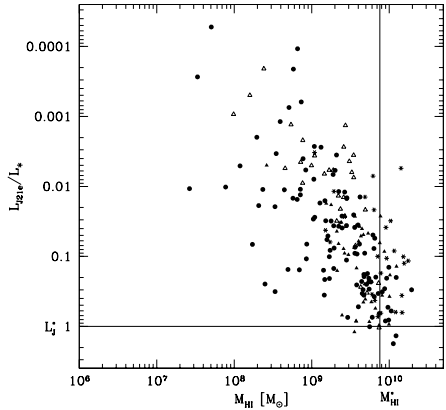}
\end{center}
\caption{The relationship between J-band luminosity and HI mass for HI-selected
galaxies. Note that there are many galaxies in the 3 blind HI surveys
represented that are missing because they were too faint or too low surface
brightness to be detected with 2MASS. The filled circles are galaxies from the
ADBS (Rosenberg \& Schneider 2000), the stars are from the AHISS survey (Zwaan
et al. 1997), and the triangles are from the Arecibo Slice survey (Spitzak \&
Schneider 1998) where the open triangles are I-band measurements to which a
color correction has been applied.}
\label{fig:massmag}
\end{figure}

Figure \ref{fig:massmag} shows the relationship between HI mass and near
infrared luminosity for 3 HI-selected galaxy samples: the ADBS (filled circles),
the Arecibo Slice survey (Spitzak \& Schneider 1998; triangles); and the 
AHISS survey (Zwaan et al. 1997; stars). The open triangles show the
Arecibo Slice galaxies for which there were I-band data, but no J-band detection
in 2MASS, so a color-corrected I-band luminosity was substituted.  
While there is a correlation between the the HI mass and the J-band luminosity, 
there is a spread of several orders of magnitude. 
This figure illustrates why we detect a large number of low luminosity systems
in absorption line studies. The DLAs are mostly drawn from an HI-rich population 
which spans the range of luminosities from a few tenths L$_\ast$ to $< 10^{-4}$
L$_\ast$. Note that many of the HI-survey sources are not in this figure because 
they were too low luminosity or too low surface brightness to be detected by 
2MASS.

\begin{figure}[ht]
\begin{center}
\epsfxsize=6in
\leavevmode
\epsffile{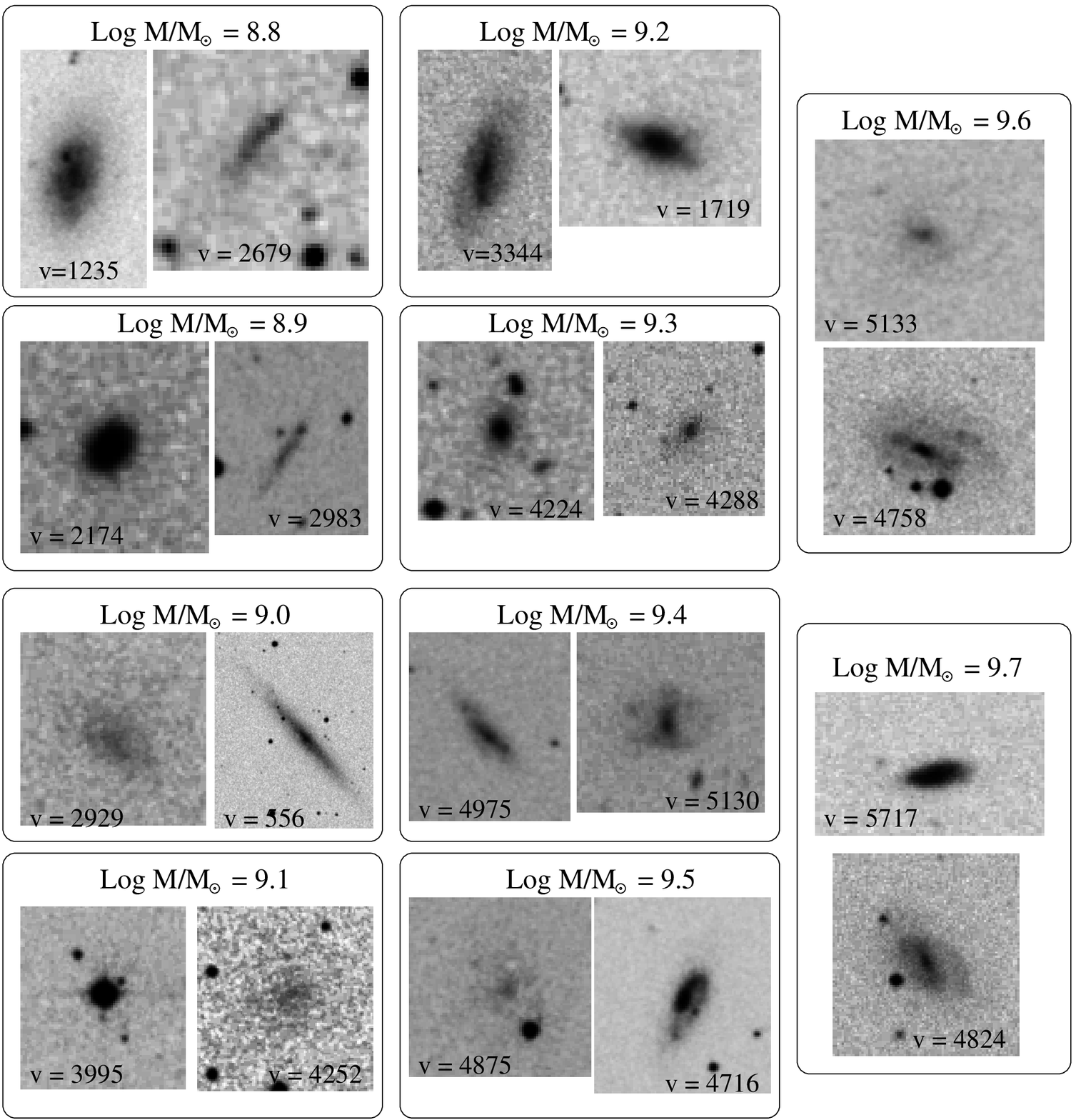}
\end{center}
\caption{Digitized Sky Survey images for a random sample
(just chosen such that there would be 2 at each mass) of ADBS galaxies with 
masses between Log(M$_{HI}/M\solar\ $) = 8.8 and 9.7. Note the large range of
morphologies, luminosities, and surface brightnesses for this sample. In fact,
there is very little correlation between these properties and the HI-mass of the
galaxy.}
\label{fig:mstar}
\end{figure}

Figure \ref{fig:mstar} shows Digitized Sky Survey images for a random sample
(just chosen such that there would be 2 at each dex) of ADBS galaxies with 
masses between Log(M$_{HI}/M\solar\ $) = 8.8 and 9.7. Note the large range of
morphologies, luminosities, and surface brightnesses for this sample. In fact,
there is very little correlation between these properties and the HI-mass of the
galaxy. This figure illustrates that the HI cross-section should not be expected
to be dominated by L$_\ast$ spirals. Many of the counterparts to DLAs should be
low surface brightness and/or irregular galaxies so care must be taken when
correlating the nearest bright spiral with a DLA at high redshift.  

\section{CONCLUSIONS}

The number density of HI-rich galaxies at low-$z$ does not require evolution of
the DLA population as indicated by Rao \& Turnshek (2000). However, the large
errorbars on the values of dN/dz at any given redshift, and the uncertainties in
the HI covering fraction, make it impossible to rule out evolution of this
population. 

The optical and kinematic properties of HI-selected galaxies are consistent with
those of DLAs. The surprising nature of the DLA population is, at least in part,
the result of an improper comparison sample rather than a detection bias. There 
may still be some subtle biases in the DLA selection, but until more 
data are available for comparing these samples, it will remain uncertain.

The galaxies which dominate the HI cross-section span a large range of
properties. They cover a large range in HI mass as well as optical luminosity,
surface brightness, and morphology. As has been found for the low redshift DLAs,
this is not a homogeneous population of bright spirals as might have been
anticipated but is, instead, a very diverse galaxy population.

\begin{acknowledgements}

J.L.R. would like to thank John Stocke for helpful discussions about this work.
The Digitized Sky Surveys were produced at the Space Telescope Science Institute
under U.S. Government grant NAG W-2166. The images of these surveys are based on
photographic data obtained using the Oschin Schmidt Telescope on Palomar
Mountain and the UK Schmidt Telescope. The plates were processed into the
present compressed digital form with the permission of these institutions.

\end{acknowledgements}

\end{document}